\documentclass[twocolumn,aps]{revtex4-1}

\pdfpageattr {/Group << /S /Transparency /I true /CS /DeviceRGB>>}

\usepackage{graphicx} 
\usepackage{overpic}

\newcommand{\ket}[1]{\left|#1\right\rangle}

\DeclareSymbolFont{lettersA}{U}{txmia}{m}{it}
\DeclareMathSymbol{\real}{\mathord}{lettersA}{"92}
\DeclareMathSymbol{\field}{\mathord}{lettersA}{"83}

\begin{document}

\title{Polyestimate: instantaneous open source surface code analysis}

\author{Austin G. Fowler}

\affiliation{Centre for Quantum Computation and Communication
Technology, School of Physics, The University of Melbourne, Victoria
3010, Australia}

\date{\today}

\begin{abstract}
The surface code is highly practical, enabling arbitrarily reliable quantum computation given a 2-D nearest-neighbor coupled array of qubits with gate error rates below approximately 1\%. We describe an open source library, Polyestimate, enabling a user with no knowledge of the surface code to specify realistic physical quantum gate error models and obtain logical error rate estimates. Functions allowing the user to specify simple depolarizing error rates for each gate have also been included. Every effort has been made to make this library user-friendly.
\end{abstract}

\maketitle

Quantum computing hardware is not expected to achieve the same level of reliability as classical computing hardware due to its complexity and reliance on the fragile phenomena of quantum mechanics. Arbitrarily reliable quantum computation can, however, be achieved through the use of quantum error correction \cite{Knil96b,Ahar97,Kita97,Fowl12e}. A bright hope in the field of quantum error correction is the surface code \cite{Brav98,Denn02,Raus07,Raus07d,Fowl12f}, which has the very experimentally reasonable requirements of a 2-D array of nearest-neighbor coupled qubits capable of implementing initialization, measurement, and one- and two-qubit unitary gates, all with error rates below approximately 1\% \cite{Wang11,Fowl11b}. Trade-offs are also possible, such as a measurement error rate of 10\% or more at the cost of somewhat lower two-qubit gate error rates. An open source analysis tool, Autotune \cite{Fowl12d}, exists that is capable of taking into account such error model details.

\begin{figure}
\begin{center}
\resizebox{60mm}{!}{\includegraphics{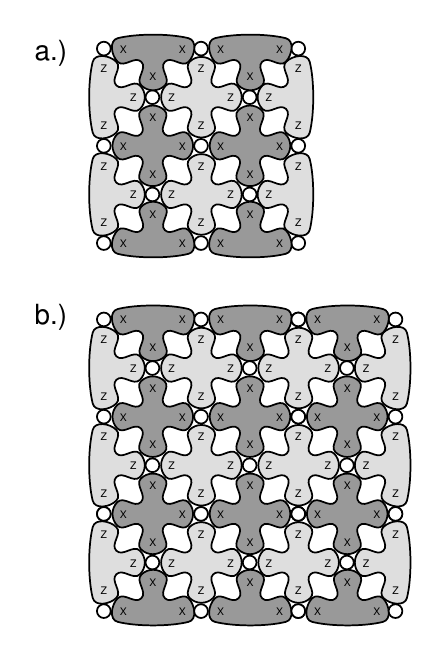}}
\end{center}
\caption{a.) Distance 3 surface code. b.) Distance 4 surface code. Circles represent qubits. Bubbles represent operators (tensor product of Pauli $X$ or $Z$ operators) that are measured to detect errors. Note that all operators commute. Each bubble is associated with its own (syndrome) qubit, used to measure its operator (stabilizer \cite{Gott97}) via the circuits shown in Fig.~\ref{syn_circs}. Each surface code contains a single logical qubit. A distance $d$ surface code, properly implemented, can correct any combination of $\lfloor(d-1)/2\rfloor$ errors.}\label{sc}
\end{figure}

\begin{figure}
\begin{center}
\resizebox{60mm}{!}{\includegraphics{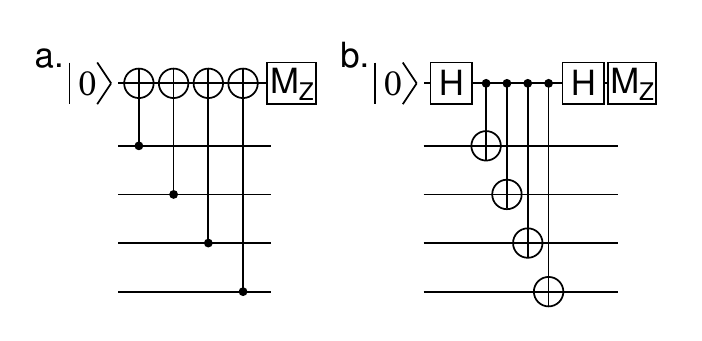}}
\end{center}
\caption{Quantum circuits measuring a.)~ZZZZ and b.)~XXXX operators. The $\ket{0}$ represents initialization, the $H$ represents Hadamard, the $M_Z$ represents measurement of the operator $Z$, and the dot and target symbols connected by lines each represent a CNOT gate. The interaction sequence is with the North data qubit as shown in Fig.~\ref{sc}, then West, East, South. If no data qubit is present in a given direction, the CNOT is simply replaced with an identity gate of duration CNOT.}\label{syn_circs}
\end{figure}

Autotune is, however, rather computationally expensive to run and complex to use. Consider Fig.~\ref{sc}. This contains two examples of different size surface codes containing a single logical qubit. The operators in Fig.~\ref{sc} are measured using the circuits of Fig.~\ref{syn_circs}. The circuits of Fig.~\ref{syn_circs} are built out of eight types of gates, namely initialization, measurement, Hadamard, CNOT, and four potentially different identity gates of durations equal to each of these nontrivial gates. It would be nice if a simple library existed taking error models or simple error rates for each gate and a code distance $d$ as input and producing accurate estimates of the logical $X$ and $Z$ error rates of the protected logical qubit. In this work, we report the creation of Polyestimate, a library which returns such estimates essentially instantaneously.

Polyestimate is designed to work accurately on error models that are not too extremely asymmetric within a single error type on the CNOT gate. For example, Polyestimate will return inaccurate results (meaning excessively high logical error rate estimates) if, for some reason, the probability of $X$ error on the control qubit of the CNOT is very much higher than the probability of an $X$ error on the target qubit. Asymmetry between $X$ and $Z$ errors is, however, accurately handled on all gates. Polyestimate can also work quite accurately quite close to threshold --- it does not rely on error rates being low as we did in \cite{Fowl12g}, work which required error rates to be of order $10^{-4}$ or lower. These statements will be made more precise as the details of Polyestimate are explained.

The discussion is organized as follows. In Section~\ref{ec}, we describe how to reduce detailed error models for the eight types of gates described above to three simple error rates. In Section~\ref{data}, we describe our simulation method and give selected examples of the data collected. In Section~\ref{interpolation}, we describe how Polyestimate interpolates and extrapolates from the simulation generated data to the user specified error models and code distance $d$. In Section~\ref{verify}, we compare the output of Polyestimate to that of detailed Autotune simulations for a range of situations, characterizing the accuracy of Polyestimate. Section~\ref{conc} concludes and discusses further work.

\section{Error classes}
\label{ec}

When using circuits of the form shown in Fig.~\ref{syn_circs}, currently the only available algorithm to correct errors is based on minimum weight perfect matching \cite{Edmo65a,Edmo65b,Fowl12c}. This algorithm currently does not take into account correlations between $X$ and $Z$ errors. For example, if after a given quantum gate there is a probability $p$ of $X$, $Y$, or $Z$ error, each with probability $p/3$, then if we believe with high confidence that we have detected an $X$ error after this gate we should also believe there to be a $Z$ error with 50\% probability since $Y=iXZ$.

From the point of view of building Polyestimate, the independent treatment of $X$ and $Z$ errors is, however, enormously useful. An arbitrary single-qubit error model with probabilities $p_X$, $p_Y$, $p_Z$ of errors $X$, $Y$, $Z$ can simply be treated as two independent error models implying an $X$ error with probability $p'_X=p_X+p_Y$ and a $Z$ error with probability $p'_Z=p_Y+p_Z$. No inaccuracy is introduced from the point of view of matching the detailed simulated logical error rate with a simple calculated logical error rate.

CNOT gate error models are not so straightforward. Given 15 error probabilities $p_{IX}$, $p_{IY}$, \ldots, $p_{ZZ}$ one can certainly make a similar identification $p'_{IX}=p_{IX}+p_{IY}+p_{ZX}+p_{ZY}$, $p'_{XX}=p_{XX}+p_{XY}+p_{YX}+p_{YY}$, etc, however if $p'_{XX}\neq p'_{IX}$ or $p'_{XX}\neq p'_{XI}$ we first artificially increase one or two of these derived error rates to obtain a balanced $p'_{IX} = p'_{XI} = p'_{XX}$ error model then define $p_{2X}=5(p'_{IX}+p'_{XI}+p'_{XX})/4$. The multiplicative factor of $5/4$ scales the 12 fundamental error rates making up $p'_{IX}$, $p'_{XI}$, $p'_{XX}$ to a 15 component depolarizing error rate with the correct probability of each of the derived $X$ terms. If the CNOT error model is very asymmetric within a single type of error, the process of raising one or two of $p'_{IX}$, $p'_{XI}$, $p'_{XX}$ can make the logical error rate estimates excessively high. We shall give a concrete example of this in Section~\ref{interpolation}. Repeating the above process for $Z$ errors gives a second value $p_{2Z}$.

Referring to Fig.~\ref{syn_circs}, data qubit identity errors can be lumped into a single variable $p_{1X}=3(p'_X({\rm IdInit})+2p'_X({\rm IdHad})+p'_X({\rm IdMeas}))/8$, and similarly for $Z$ errors, giving $p_{1Z}$. The abbreviations stand for identity of duration initialization, identity of duration Hadamard, and identity of duration measurement, respectively. Identity of duration CNOT does not appear in the equation as in even a small distance surface code only a few such identity gates are required around the boundary of the qubit lattice and their influence is negligible --- the vast majority of data qubits are busy performing CNOTs when any data qubit is involved in a CNOT. While this may not be immediately apparent from Fig.~\ref{syn_circs}, it should be remembered that all stabilizers are measured simultaneously and each data qubit not on the boundary of the surface code interacts with each of its four neighboring syndrome qubits. The factor of $3/8$ in the definition of $p_{1X}$ consists of a factor of $3/2$ to scale the individual error type error rate to a depolarizing error rate, and a factor of $1/4$ to reduce the result to an error rate per gate rather than per stabilizer measurement cycle.

Syndrome qubits associated with $Z$ stabilizer measurements, meaning $X$ error detection, only suffer initialization and measurement errors in addition to CNOT errors. Initialization and measurement errors cannot propagate off the syndrome qubit. We can therefore define $p_{0X}=p_X({\rm Init})+p_X({\rm Meas})$. Note that for initialization and measurement $p_Y=p_Z=0$. For syndrome qubits associated with $X$ stabilizer measurements, meaning $Z$ error detection, we have two additional Hadamard gates, the first of which is only dangerous if it introduces $Z$ errors, the second of which is only dangerous if it introduces $X$ errors. We therefore get a second definition $p_{0Z}=p_X({\rm Init})+p'_X({\rm Had})+p'_Z({\rm Had})+p_X({\rm Meas})$. We choose to leave $p_{0A}$ as an error rate per error detection cycle as the number of gates for $A=X$ and $A=Z$ differs.

Given arbitrary error models, we have described how to obtain two sets of error rates $p_{0A}$, $p_{1A}$, $p_{2A}$ focusing on either $A=X$ or $A=Z$ errors. These simple error rates well characterize the behavior of the detailed error models provided the CNOT error model does not deviate very far from balanced depolarizing within a single type of error.

\section{Simulations}
\label{data}

It is straightforward to use Autotune to simulate the performance of a given surface code of distance $d$ making use of gate error rates $p_{0A}$, $p_{1A}$, $p_{2A}$. Autotune will set up the appropriate array of qubits, execute the appropriate sequence of gates, insert stochastic errors and use minimum weight perfect matching to insert corrections. When too many errors occur in the wrong places, logical errors will occur. Autotune simulates many continuous cycles of error detection and calculates the probability of logical error per round.

Our goal is to obtain the logical $X$ and $Z$ error rates corresponding to a sufficiently broad range of $d$, $p_{0A}$, $p_{1A}$, $p_{2A}$ to enable the logical $X$ and $Z$ error rates at any other combination of values of $d$, $p_{0A}$, $p_{1A}$, $p_{2A}$ to be accurately determined.

Note that below threshold the logical error rate is provably exponentially suppressed with increasing $d$ \cite{Fowl12e}, or, to be more precise, exponentially suppressed with increasing $d_e=\lfloor (d+1)/2\rfloor$. This means that we only need to simulate distances $d=3,4,5,6$ as the logical error rate at all higher distances can be obtained by taking the ratio of odd or even distance data to the appropriate power with the appropriate pre-factor. Details are given in Section~\ref{interpolation}.

Note that the threshold error rate when the surface code is subjected to only CNOT gate errors is approximately 1.25\% \cite{Fowl12f}. We therefore do not simulate values of $p_{2A}$ above 2\%, as this is sufficiently high to be above threshold for all values of $p_{0A}$ and $p_{1A}$.

Let $r_{0A}=p_{0A}/p_{2A}$ and $r_{1A}=p_{1A}/p_{2A}$. We have initially considered only syndrome errors in the range $r_{0A}\in [0.01,200]$ and data qubit errors in the range $r_{1A}\in [0.01,1]$. This is essentially a statement that we consider it possible that initialization and measurement error rates could be significantly less, comparable, or significantly greater than the CNOT error rate, whereas we expect the single-qubit identity error rates to be less than or equal to the CNOT error rate. It would be straightforward to simulate a broader range if a quantum technology called for it. We have restricted ourselves to $p_{2A}\geq 10^{-4}$ as no scalable quantum technology has even achieved a CNOT gate with error rate $10^{-2}$, and hence the range $p_{2A}\in [10^{-4},0.02]$ covers all existing technology, and technology likely to be available in the short to medium term. To keep the size of the simulation task under control, we have initially restricted ourselves to three points per decade, so for example $r_{1A}\in \{0.01,0.02,0.05,0.1,0.2,0.5,1\}$.

\begin{figure}
\begin{center}
\resizebox{85mm}{!}{\includegraphics[viewport=60 60 545 430, clip=true]{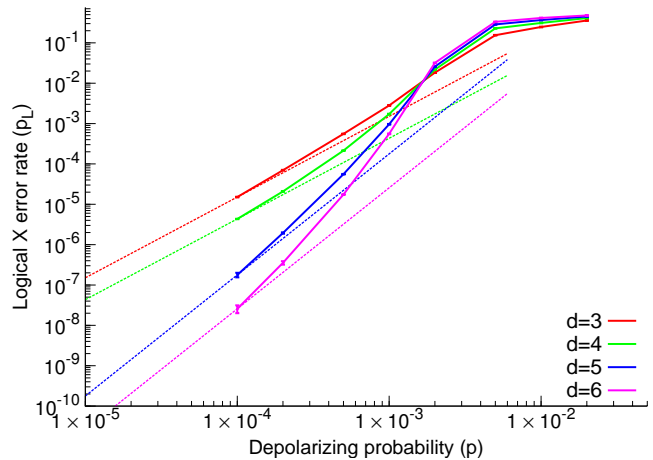}}
\end{center}
\caption{(Color online) probability of logical $X$ error as a function of the depolarizing error rate $p=p_{2A}$ for distances $d=3,4,5,6$ and $r_{0A}=100$, $r_{1A}=1$. Referring to the left of the figure, the distance $d=3,4,5,6$ curves are ordered top to bottom. Quadratic curves have been fit through the lowest $d=3,4$ data points, and cubic curves through the lowest $d=5,6$ data points.}\label{logx_100_1}
\end{figure}

\begin{figure}
\begin{center}
\resizebox{85mm}{!}{\includegraphics[viewport=55 60 545 430, clip=true]{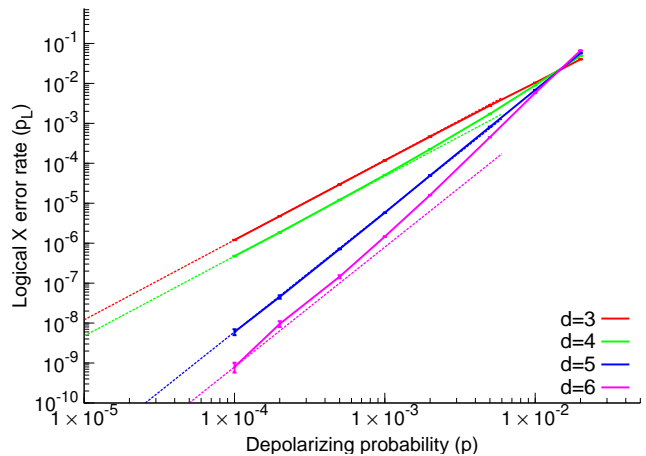}}
\end{center}
\caption{(Color online) probability of logical $X$ error as a function of the depolarizing error rate $p=p_{2A}$ for distances $d=3,4,5,6$ and $r_{0A}=0.01$, $r_{1A}=0.01$. Referring to the left of the figure, the distance $d=3,4,5,6$ curves are ordered top to bottom. Quadratic curves have been fit through the lowest $d=3,4$ data points, and cubic curves through the lowest $d=5,6$ data points.}\label{logx_0.01_0.01}
\end{figure}

Samples of the data collected can be found in Figs.~\ref{logx_100_1}--\ref{logx_0.01_0.01}. These two graphs contain data for the highest and lowest error ratios, respectively. Error bars corresponding to $2\sigma$ confidence have been included. At the very lowest error rates $p_{2A}$ it is computationally challenging to observe a sufficiently large number of logical errors to obtain an accurate logical error rate. Nevertheless, the data presented is sufficiently accurate for practical purposes. This data is available as part of the Polyestimate library, available online at \cite{Fowl12d}.

\section{Interpolation and extrapolation}
\label{interpolation}

Given detailed error models for each gate or simple error rates for each gate, in Section~\ref{ec} we described how to define two sets of three values $r_{0A}=p_{0A}/p_{2A}$, $r_{1A}=p_{1A}/p_{2A}$, $p_{2A}$, where $A=X,Z$, that can be used to estimate the failure behavior of the surface code. In Section~\ref{data}, we described the construction of a database of logical error rates for a range of code distances and values of $r_{0A}$, $r_{1A}$, $p_{2A}$. Note that a single database suffices for both $A=X$ and $A=Z$ as the database is for balanced depolarizing noise. We now describe how to obtain functions of $d$ giving estimates of the logical $X$ and $Z$ error rates.

If $r_{0A}, r_{1A},p_{2A}\neq 1,2,5\times 10^{i}$ for some integer i, interpolation will be required. Define superscripts $H$ and $L$ to mean the first variable value in the database higher or lower than the given value. For example, if $r_{0A}=0.3$, then $r^L_{0A}=0.2$ and $r^H_{0A}=0.5$. Note that the logical error rates are strictly increasing smooth functions of $r_{0A}$, $r_{1A}$, $p_{2A}$. For a sufficiently fine-grained database, linear interpolation is therefore appropriate.

For each distance $d=3,4,5,6$ in the database and each type of logical error $A$, find the up to eight different values of logical $A$ error for the up to eight different combinations of $r^B_{0A}$, $r^B_{1A}$, $p^B_{2A}$, where $B=H,L$. Linearly interpolate and define this to be the logical error rate $p_{A_L}(d,r_{0A},r_{1A},p_{2A})$. For brevity, we shall refer to this as simply $p_{A_L}(d)$.

For $d>6$, we extrapolate. Define $x=p_{A_L}(5)/p_{A_L}(3)$, $C=p_{A_L}(3)/x^2$, $y=p_{A_L}(6)/p_{A_L}(4)$, $D=p_{A_L}(4)/y^2$. This gives us four expressions
\begin{eqnarray}
p_{A_L}(d) & = & Cx^{\lfloor \frac{d+1}{2} \rfloor}, \label{odd} \\
p_{A_L}(d) & = & Dy^{\lfloor \frac{d+1}{2} \rfloor}, \label{even}
\end{eqnarray}
valid for $d$ odd and even, respectively, and $A=X,Z$.

\section{Verification}
\label{verify}

The simplicity of derivation and form of Eqs.~\ref{odd}--\ref{even} is appealing, however verification of the accuracy of logical error rates obtained by Polyestimate is required. In this Section, we shall compare the output of Polyestimate with data generated in a number of situations by direct simulation with Autotune.

The first situation we shall consider is balanced depolarizing noise of equal strength on all gates. When all gates have probability $p=10^{-3}$ of failure, Table~\ref{comp} shows in columns 3 and 4 the Autotune determined probability of logical $X$ and $Z$ error for distances $d=3,\ldots,8$. This should be compared with the Polyestimate data in columns 5 and 6. As the distance increases, the computational cost of using Autotune rapidly becomes prohibitive, whereas Polyestimate can be used for arbitrary distances, and generated the data in Table~\ref{comp} in approximately one hundredth of one second, including loading the database from disk and calculating the appropriate values in Eqs.~\ref{odd}--\ref{even}. The computational cost of evaluating Eqs.~\ref{odd}--\ref{even} for additional values of $d$ is negligible after loading the database.

\begin{table}
\begin{tabular}{c|c|c|c|c|c}
$d$ & $T$ & $p_{X_L}$ & $p_{Z_L}$ & $p_{X_L}(d)$ & $p_{Z_L}(d)$ \\
\hline
3 & 10min & $1.1\times 10^{-3}$ & $1.4\times 10^{-3}$ & $1.1\times 10^{-3}$ & $1.4\times 10^{-3}$ \\
4 & 90min & $4.5\times 10^{-4}$ & $5.8\times 10^{-4}$ & $4.5\times 10^{-4}$ & $5.8\times 10^{-4}$ \\
5 & 7.7hr & $1.0\times 10^{-4}$ & $1.5\times 10^{-4}$ & $9.9\times 10^{-5}$ & $1.4\times 10^{-4}$ \\
6 & 48hr & $3.2\times 10^{-5}$ & $4.7\times 10^{-5}$ & $3.2\times 10^{-5}$ & $4.7\times 10^{-5}$ \\
7 & 250hr & $8.5\times 10^{-6}$ & $1.4\times 10^{-5}$ & $9.0\times 10^{-6}$ & $1.5\times 10^{-6}$ \\
8 & 500hr & $2.5\times 10^{-6}$ & $4.2\times 10^{-6}$ & $2.2\times 10^{-6}$ & $3.8\times 10^{-6}$ \\
9 & - & - & - & $8.2\times 10^{-7}$ & $1.6\times 10^{-6}$ \\
10 & - & - & - & $1.6\times 10^{-7}$ & $3.0\times 10^{-7}$ \\
36 & - & - & - & $1.5\times 10^{-22}$ & $1.7\times 10^{-21}$ \\
\end{tabular}
\caption{Comparison of Autotune and Polyestimate for depolarizing noise of rate $p=10^{-3}$. Column 1 contains the code distance $d$. Column 2 contains the total CPU time required by Autotune to generate the logical $X$ and $Z$ error rates shown in columns 3 and 4. Columns 5 and 6 contain Polyestimate generated logical $X$ and $Z$ error rates. Agreement is in all cases within 10\%, sufficient for practical purposes.}
\label{comp}
\end{table}

\begin{table}
\begin{tabular}{c|c|c|c|c|c}
$d$ & $T$ & $p_{X_L}$ & $p_{Z_L}$ & $p_{X_L}(d)$ & $p_{Z_L}(d)$ \\
\hline
3 & 11min & $2.8\times 10^{-3}$ & $3.4\times 10^{-3}$ & $2.8\times 10^{-3}$ & $3.4\times 10^{-3}$ \\
4 & 40min & $1.8\times 10^{-3}$ & $2.2\times 10^{-3}$ & $1.7\times 10^{-3}$ & $2.3\times 10^{-3}$ \\
5 & 2hr & $9.6\times 10^{-4}$ & $1.3\times 10^{-3}$ & $9.9\times 10^{-4}$ & $1.3\times 10^{-3}$ \\
6 & 6.3hr & $5.7\times 10^{-4}$ & $7.9\times 10^{-4}$ & $5.9\times 10^{-4}$ & $8.6\times 10^{-4}$ \\
7 & 16hr & $3.4\times 10^{-4}$ & $4.9\times 10^{-4}$ & $3.5\times 10^{-4}$ & $5.0\times 10^{-4}$ \\
8 & 47hr & $2.0\times 10^{-4}$ & $3.0\times 10^{-4}$ & $2.0\times 10^{-4}$ & $3.2\times 10^{-4}$ \\
9 & 86hr & $1.2\times 10^{-4}$ & $1.9\times 10^{-4}$ & $1.2\times 10^{-4}$ & $1.9\times 10^{-4}$ \\
10 & 199hr & $7.6\times 10^{-5}$ & $1.2\times 10^{-4}$ & $6.6\times 10^{-5}$ & $1.2\times 10^{-4}$ \\
11 & 422hr & $4.6\times 10^{-5}$ & $7.8\times 10^{-5}$ & $4.3\times 10^{-5}$ & $7.5\times 10^{-5}$ \\
12 & 960hr & $2.8\times 10^{-5}$ & $4.8\times 10^{-5}$ & $2.2\times 10^{-5}$ & $4.6\times 10^{-5}$ \\
13 & 1590hr & $1.7\times 10^{-5}$ & $3.2\times 10^{-5}$ & $1.5\times 10^{-5}$ & $2.9\times 10^{-5}$ \\
14 & - & - & - & $7.4\times 10^{-6}$ & $1.7\times 10^{-5}$ \\
15 & - & - & - & $5.2\times 10^{-6}$ & $1.1\times 10^{-5}$ \\
\end{tabular}
\caption{Comparison of Autotune and Polyestimate for 10\% measurement error and depolarizing noise of rate $p=10^{-3}$ on all other gates. Column 1 contains the code distance $d$. Column 2 contains the total CPU time required by Autotune to generate the logical $X$ and $Z$ error rates shown in columns 3 and 4. Columns 5 and 6 contain Polyestimate generated logical $X$ and $Z$ error rates. Agreement is in all cases within 15\%, sufficient for practical purposes.}
\label{comp}
\end{table}

\begin{table}
\begin{tabular}{c|c|c|c}
$d$ & $T$ & $p_{X_L}$ & $p_{X_L}(d)$ \\
\hline
3 & 10min & $1.1\times 10^{-3}$ & $4.0\times 10^{-3}$ \\
4 & 47min & $4.6\times 10^{-4}$ & $2.5\times 10^{-3}$ \\
5 & 6.8hr & $9.7\times 10^{-5}$ & $1.4\times 10^{-3}$ \\
6 & 33hr & $3.1\times 10^{-5}$ & $8.7\times 10^{-4}$ \\
7 & 199hr & $7.9\times 10^{-6}$ & $4.7\times 10^{-4}$ \\
8 & 986hr & $2.3\times 10^{-6}$ & $3.0\times 10^{-4}$ \\
\end{tabular}
\caption{Comparison of Autotune and Polyestimate for the case of all gates having total probability of error $p=10^{-3}$, however the CNOT gate having an asymmetric error model satisfying $p'_{IX}=10p'_{XI}=100p'_{XX}$. Column 1 contains the code distance $d$. Column 2 contains the total CPU time required by Autotune to generate the logical $X$ error rates shown in column 3. Column 4 contains Polyestimate generated logical $X$ error rates. Agreement is poor, with Polyestimate significantly overestimating the logical error rate due to its need to raise the probability of $XI$ and $XX$ errors.}
\label{comp}
\end{table}

The agreement between Autotune and Polyestimate is within 10\% for the data shown in Table~\ref{comp}. It should be noted, however, that since Polyestimate uses a simple exponential expression (Eqs.~\ref{odd}--\ref{even}), any small inaccuracy will grow with code distance, so absolute values of logical error rates for large distances will only be approximate. A user can still, however, quite accurately determine the code distance $d$ required to suppress logical errors below some desired rate. For example, if a logical error rate below $10^{-20}$ is desired, Polyestimate can be used to predict that a code distance of approximately 36 is sufficient.

The next case we shall consider is depolarizing noise of rate $p=10^{-3}$ for every gate except the measurement gate which shall have an error rate of 10\%. It is possible that superconducting qubits may possess such a distribution of error rates as fast, low error measurement is especially challenging \cite{Chen12}. Agreement between Autotune and Polyestimate is again very good, within 15\% in all cases, frequently significantly better, and this agreement is exhibited over logical error rates varying by over two orders of magnitude. Similar levels of agreement have been observed for other ratios of simple depolarizing error rates on each gate.

Finally, we shall consider the case of all gates having total probability of error $p=10^{-3}$, however the CNOT gate having an asymmetric error model satisfying $p'_{IX}=10p'_{XI}=100p'_{XX}$. We shall focus only on logical $X$ errors. In this instance, since Polyestimate must raise the probability of $XI$ and $XX$ errors significantly in order to use its pre-generated database, the logical $X$ error estimates are excessively high. When the CNOT error model is very asymmetric within a single type of error, direct simulation using Autotune is currently the only option unless all error rates are quite low and asymptotic techniques can be used \cite{Fowl12g}.

\section{Conclusion}
\label{conc}

We have described an open source tool, Polyestimate, that is shipped with a database of Autotune simulation results sufficiently broad and detailed to provide, through interpolation and extrapolation, instantaneous surface code logical error rates for a very broad range of practical depolarizing error rates and arbitrary code distances $d$. Individual depolarizing error rates for initialization, measurement, Hadamard, CNOT, and identity gates of duration each of these gates, respectively, can be specified and handled accurately. The primary limitation of Polyestimate is its inability to cope with very asymmetric error models on the CNOT within a single type of error. In future work, we plan address this limitation with slower analytic techniques for such error models that will still be orders of magnitude faster than running Autotune.

\section{Acknowledgements}
\label{ack}

This research was conducted by the Australian Research Council Centre of Excellence for Quantum Computation and Communication Technology (project number CE110001027), with support from the US National Security Agency and the US Army Research Office under contract number W911NF-13-1-0024. Supported by the Intelligence Advanced Research Projects Activity (IARPA) via Department of Interior National Business Center contract number D11PC20166.  The U.S. Government is authorized to reproduce and distribute reprints for Governmental purposes notwithstanding any copyright annotation thereon.  Disclaimer: The views and conclusions contained herein are those of the authors and should not be interpreted as necessarily representing the official policies or endorsements, either expressed or implied, of IARPA, DoI/NBC, or the U.S. Government.

\bibliography{../References}

\end{document}